\documentstyle[aps]{revtex}
\begin{document}
\input{psfig}
\draft
\title{\bf Recovering coherence from decoherence:
a method of quantum state reconstruction}
\author{H. Moya-Cessa\footnote{Permanent address: INAOE, Coordinaci\'on de Optica, 
Apdo. Postal 51 y 216, 72000 Puebla, Pue., Mexico.}
\footnote{Email address: hmmc@inaoep.mx}
, J.A. Roversi\footnote{Email address: roversi@ifi.unicamp.br}, 
S.M. Dutra\footnote{Email address: dutra@rulhm1.leidenuniv.nl}, 
and A. Vidiella-Barranco \footnote{Email address: vidiella@ifi.unicamp.br}}
\address{Instituto de F\'\i sica ``Gleb Wataghin'',
Universidade Estadual de Campinas,
13083-970   Campinas  SP  Brazil}
\date{\today}
%\twocolumn
\maketitle
\begin{abstract}
We present a feasible scheme for reconstructing the quantum state of a 
field prepared inside a lossy cavity. Quantum coherences are normally destroyed 
by dissipation, but we show that at zero temperature we are able to retrieve 
enough information about the initial state, making possible to recover its 
Wigner function as well as other quasiprobabilities. We provide a numerical simulation 
of a Schr\"odinger cat state reconstruction.
\end{abstract}

\section{Introduction}

The reconstruction of quantum states is a central topic in quantum optics and
related fields \cite{risk,ulf}. During the past years, several techniques have been
developed, for instance, the direct sampling of the density matrix of a signal
mode in multiport optical homodyne tomography \cite {zuc}, tomographic 
reconstruction by unbalanced homodyning \cite {wal}, cascaded homodyning \cite{kis}
and reconstruction via photocounting \cite{ban}. 
There are also proposals of measurement of the 
electromagnetic field inside a cavity \cite{dav} as well as the vibrational state 
of an ion in a trap \cite {bar}. The full reconstruction of nonclassical
field states \cite{smi} as well as of (motional) states of an ion \cite{win} 
have been already experimentally accomplished. The quantum state reconstruction is 
normally achieved through a finite set of either field homodyne measurements, or 
selective measurement of atomic states \cite{dav} in the case of cavities. This makes 
possible to construct a quasidistribution (such as the the Wigner function) which 
constitutes an alternative representation of the quantum state of the field. 

Nevertheless, in real experiments, the presence of noise and dissipation has
normally destructive effects. In fact, as it has been already pointed out, the
reconstruction schemes themselves also indicate loss of coherence in quantum 
systems \cite{win}. regarding this subject, a scheme for compensation of losses in 
quantum-state measurements has been already proposed \cite{ulf1}, and 
the relation between losses and $s$-parameterized quasiprobability distributions 
has been already pointed out in \cite{ulf2}. The scheme on loss
compensation in \cite{ulf1} applies to photodetector losses, and consists 
essentially of a mathematical inversion formula expressing the initial density matrix 
in terms of the decayed one. Our scheme, as discussed in \cite{ours}, involves a
physical process that actually enables us to store information about all
the quantum coherences of the initial state in the diagonal elements
(photon distribuition) of the density matrix of a transformed state. By
storing this information in the diagonal elements, it becomes much more
robust under dissipation, allowing us to recover the Wigner function of the initial
field state in a time scale of
the order of the energy decay time that is, of course, much longer than
the extremely fast decoherence time scale that is normally associated 
with the dissipation of quantum coherences. 

We consider a single mode high-$Q$ cavity where we suppose that a (nonclassical) 
field state $\hat{\rho}(0)$ is previously prepared. The first step of our
method consists in driving the generated quantum state by a coherent pulse. 
The reconstruction of the field state may be accomplished after turning-off the 
driving field, i.e., at a time in which the cavity
field has already suffered decay. We use the fact that by displacing the 
initial state (even while it is decaying) we make its quantum coherences 
robust enough to allow its experimental determination, at a later time, despite 
dissipation. We then show that the evolution of the cavity field is such that it 
directly yields the Wigner function of the initial nonclassical field simply by
measuring the photon number distribution of the displaced field. 
For that we make direct use of the series representation of 
quasiprobability distributions \cite{moy}. A numerical simulation of our method is 
presented, and we take into account the action of dissipation while driving the initial
field.

This manuscript is organized as follows: in Sec. II we discuss, taking into 
account losses, the process of displacement of the initial field. 
In Sec. III we show how to reconstruct the initial cavity field after allowing 
the displaced field to decay. In Sec. IV we present a simulation of the
reconstruction of a Schr\"odinger cat state. In Sec. V we summarize our 
conclusions.

\section{Driving the initial field}

We assume that the initial nonclassical field $\hat{\rho}(0)$ is prepared inside a
high $Q$ cavity. The master equation in the interaction picture for the reduced density 
operator $\hat{\rho}$ relative to a driven cavity mode, taking into account cavity 
losses at zero temperature and under the Born-Markov approximation is given by 
\cite {loui}

\begin{eqnarray}
         \frac{ \partial \hat{\rho}}{\partial t} = -\frac{i}{\hbar} 
         [\hat{H}_d,\hat{\rho}] +
         \frac{\gamma}{2} \left(
           2\hat{a} \hat{\rho} \hat{a}^\dagger 
             - \hat{a}^\dagger \hat{a} \hat{\rho} - \hat{\rho} \hat{a}^\dagger
             \hat{a}\right),    \label{1}
\end{eqnarray} 
with 
\begin{equation}
\hat{H}_d = i \hbar\left(\alpha ^*\hat{a}-\alpha \hat{a}^\dagger\right),
\end{equation}    
where $\hat{a}$ and $\hat{a}^\dagger$ are the annihilation and
creation operators, $\gamma$ the (cavity) decay constant and $\alpha$
the amplitude of the driving field. 

We define the superoperators $\hat{\cal R}$ and $\hat{\cal L}$ by their action 
on the density operator \cite{6}

\begin{equation}
\hat{\cal R}\hat{\rho} = (\alpha^*\hat{a}-\alpha \hat{a}^\dagger)\hat{\rho} -
                    \hat{\rho}(\alpha^*\hat{a}- \alpha \hat{a}^\dagger),\label{2}
\end{equation}
and
\begin{equation}                   
\hat{\cal L}\hat{\rho} = \gamma \hat{a}\hat{\rho}\hat{a}^\dagger
                 -\frac{\gamma}{2} \left(\hat{a}^\dagger \hat{a} \hat{\rho} +
\hat{\rho} 
\hat{a}^\dagger\hat{a}\right).\label{3}
\end{equation}
It is not difficult to show that
\begin{equation}
[\hat{\cal R},\hat {\cal L}]\hat{\rho} = \frac{\gamma}{2} \hat{\cal R}\hat{\rho},
\end{equation}
and the formal solution of Eq. (\ref{1}) can then be written as \cite {7}

\begin{equation}
\hat{\rho}(t) = \exp\left[(\hat{\cal R}+\hat{\cal L})t\right]\hat{\rho}(0) 
        =\exp(\hat{\cal L}t)\exp\left[-\frac{2\hat{\cal R}}{\gamma}
          (1-e^{\gamma t/2})\right]\hat{\rho}(0). \label{4}
\end{equation}
After driving the initial field during a time $t_d$, the resulting field density
operator will read
\begin{equation}
\hat{\rho}(t_d) = e^{\hat{\cal L}t_d} \hat{\rho}_{\beta}(0),
\end{equation}
where
\begin{equation}
\hat{\rho}_{\beta}(0) = \hat{D}^\dagger(\beta)\hat{\rho}(0)\hat{D}(\beta),
\label{5}
\end{equation}
and with
\begin{equation}
\beta = 2\alpha \frac{1-e^{\gamma t_d/2}}{\gamma}.
\label{7}
\end{equation}

This means that if we drive the initial field while it decays, during a time 
$t_d$, this is equivalent to having the field driven by a coherent field with an 
effective amplitude $\beta$ given in Eq. (\ref{7}).

\section{The reconstruction method}

The driving of the initial field is carried out during a time $t_d$. 
This procedure will enable us to
obtain information about all the elements of the initial density
matrix from the diagonal elements of the time-evolved displaced density matrix
only. As diagonal elements decay much slower than off-diagonal ones, 
information about the initial state stored this way becomes robust enough to 
withstand the decoherence process. We will now show how this robustness can be
used to obtain the Wigner function of the initial state after it has already 
started to dissipate. Once the injection of the coherent pulse is completed, 
the cavity field is left to decay, so that its dynamics will be governed by the 
master
equation in Eq. (\ref{1}) without the first (driving) term in its right-hand-side.
Therefore, the cavity field density operator will be, at a time $t$, given by

\begin{equation}
\hat{\rho}_\beta(t) = e^{(\hat{J}+\hat{L})t}\hat{\rho}_\beta (0) ,
\label{8}
\end{equation} 
with

\begin{equation}
\hat{J}\hat{\rho} = \gamma \hat{a}\hat{\rho}\hat{a}^\dagger, \ \ \ \ 
\hat{L}\hat{\rho} = -\frac{\gamma}{2} \left(\hat{a}^\dagger \hat{a} \hat{\rho} +
\hat{\rho} 
\hat{a}^\dagger\hat{a}\right).    \label{9}
\end{equation}

The next step is to calculate the diagonal matrix elements of 
$\hat{\rho}_\beta (t)=\exp\left[(\hat{J}+\hat{L})t\right]\hat{\rho}_\beta$ 
in the number state basis, or

\begin{equation}
\langle m|\hat{\rho}_\beta (t)|m\rangle=\frac{e^{-m\gamma t}}{q^m}\sum_{n=0}^{\infty}
q^n \left(\begin{array}{c} n \\ m \end{array} \right)\langle n|
\hat{\rho}_\beta (0) |n\rangle,
\end{equation}
where $q=1-e^{-\gamma t}$.

Now we multiply those matrix elements by powers of the function 

\begin{equation}
\chi(t)=1-2e^{\gamma t}. \label{chi}
\end{equation}

If we sum the resulting expression over $m$, we obtain the following simple sum

\begin{equation}
F_W=\frac{2}{\pi}\sum_{m=0}^\infty \chi^m(t)\langle m|\hat{\rho}_\beta (t)|m\rangle=
\frac{2}{\pi}\sum_{n=0}^\infty (-1)^n\langle n|\hat{D}^{\dagger}(\beta)\hat{\rho}(0)
\hat{D}(\beta)|n\rangle.\label{trans}
\end{equation}

The expression in Eq. (\ref{trans}) is exactly the Wigner function corresponding to 
$\hat{\rho}$ (the initial field state) \cite{moy} at the point specified by the
complex amplitude $\beta$. Therefore we need simply to measure the diagonal 
elements of the dissipated displaced cavity field 
$P_m(\beta;t)=\langle m|\hat{\rho}_\beta (t)|m\rangle$ for a range of $\beta$'s, 
the transformation in Eq. (\ref{trans}) in order to obtain the Wigner function 
of the initial state for this range. We note that after performing the sum, the
time-dependence cancels out completely, leaving us a constant Wigner function, as
it should be. Therefore the initial state may be reconstructed, at least in
principle, at an arbitrary later time. In practice, however, the decay of the field 
energy will impose a limitation on the times during which we will be able to measure 
the $P_m$s.  

The next step in our scheme is to measure the field photon number distribution 
$P_m(\beta;t)$. 
In a cavity, particularly in the microwave regime where there are no photodetectors 
available, experimentalists have been forced to use atoms instead to 
probe the intra-cavity field. One way of determining $P_m$ is by injecting 
atoms into the cavity and measuring their population inversion as they exit
after an interaction time $\tau$ much shorter than the cavity decay time.
It is convenient in this case to use three-level atoms in a cascade configuration with 
the upper and the lower level having the same parity and satisfying the two-photon 
resonance condition. The population inversion in this case is \cite{dan}

\begin{equation}
W(\alpha;t+\tau)=\sum_{n=0}^{\infty}P_n(\alpha;t)\left[\frac{\Gamma_n}{\delta^2_n}+
\frac{(n+1)(n+2)}{\delta^2_n}
\cos\left(2\delta_n\lambda\tau\right)\right],\label{inv}
\end{equation}
where $\Gamma_n=\left[\Delta+\chi(n+1)\right]/2$, $\delta_n^2=\Gamma_n^2+
\lambda^2(n+1)(n+2)$,
$\Delta$ is the atom-field detuning, $\chi$ is the Stark shift coefficient, and 
$\lambda$ is the atom-field coupling constant. Now we take $\Delta=0$ (two-photon 
resonance condition) and $\chi=0$. It is a good approximation to make
$\left[(n+1)(n+2)\right]^{1/2}\approx n+3/2$, so that the population inversion 
reduces to
\begin{equation}
W(\alpha;t+\tau)=\sum_{n=0}^{\infty}P_n(\alpha;t)
\cos\left(\left[2n+3\right]\lambda\tau\right).\label{invs}
\end{equation}   
This represents the atomic response to the displaced field. In order to obtain
$P_m$ from a family of measured population inversions, we need to invert the Fourier 
series in Eq. (\ref{invs}), or 
\begin{equation}
P_m(\beta;t)=\frac{2\lambda}{\pi}\int_0^{\tau_{max}} d\tau\:W(t+\tau)
\cos\left(\left[2m+3\right]\lambda\tau\right). \label{pn}
\end{equation}
We need a maximum interaction time $\tau_{max}=\pi/\lambda$  
much shorter than the cavity decay time, and this condition implies that we must
be in the strong-coupling regime, i.e. $\lambda\gg\gamma$. 
Of course after some time the atoms sent through the cavity will not be able 
to respond anymore to the decaying field, hampering our reconstruction scheme. 
Nevertheless this happens at a time scale of $t_d\approx 1/\lambda$, which is
normally much longer than a typical decoherence time.

Our scheme is easily extended to other ($s$-parametrized \cite{9}) 
quasi-probability distributions, which also may be expressed as a series \cite{moy}, 
\begin{equation}
F(\beta;s)=-\frac{2}{\pi(s-1)}\sum_{n=0}^\infty \left(\frac{s+1}{s-1}\right)^n
\langle n|\hat{\rho}_\beta |n\rangle.
\end{equation}
For this we have to multiply the photon number distribution of the displaced 
(and dissipated) field by a generalization of the function in Eq. (\ref{chi}), or
\begin{equation}
\chi(s;t)=1+\frac{2e^{\gamma t}}{s-1} \label{chig}.
\end{equation}
This increases our possibilities of measuring quantum states. We
have available several quasiprobabilities, and we may choose a more convenient
function depending on the particular conditions of a reconstruction experiment.

\section{Reconstruction of a Schr\"odinger cat state}

We show now how our method may be applied to a specific case, e.g., to a
reconstruction of a
Schr\"odinger cat state, represented by a quantum superposition of two coherent 
states having distinct amplitudes, $|\alpha\rangle$ and $|-\alpha\rangle$. The
density operator corresponding to such state is
\begin{equation}
\hat{\rho}(0)={\cal N}\left[|\alpha\rangle\langle\alpha|+
|-\alpha\rangle\langle -\alpha|+e^{i\phi}|\alpha\rangle\langle -\alpha|+
e^{-i\phi}|-\alpha\rangle\langle\alpha|\right] \label{scat},
\end{equation}
where $\phi$ is a relative phase and ${\cal N}$ a normalization constant.
Schr\"odinger cat states are very fragile under dissipation \cite{vid}, and
therefore are specially suitable for exemplifying our scheme.

In a real experiment, just after having prepared the field in the state in Eq. 
(\ref{scat}) by one of the conventional methods, (see reference \cite{haro} for
instance), the cavity will be driven by a coherent field, say, of amplitude 
$\alpha_0$. After that, at a time $t_0$, $N$ conveniently prepared three-level 
atoms should cross the cavity, so that we are able to assign a particular 
population inversion $W_{\alpha_0}(t_0)=P_e(t_0)-P_g(t_0)$ to that time. These
sort of
measurements for the {\it same\/} driving field amplitude $\alpha_0$ will be
repeated for a range of times $t_0,\; t_1, \; t_2\;\dots \; t=t_0+\pi/\lambda$, so 
that we obtain a set of values for the population inversion in that time interval. 
Now through numerical integration (see Eq. (\ref{pn})), we obtain the required 
photon number distribution $P_m(\alpha_0,t)$. This is exactly what we need in our
reconstruction scheme. The next step is to multiply $P_m$ by the terms
$\chi(t)^m=(1-2e^{\gamma t})^m$ and sum over $m$. 
This directly yields, according to our scheme, the value of the Wigner 
function of the initial field in Eq. (\ref{scat}) at the point $\alpha_0=x_0+iy_0$
in phase space.
We remark that the convergence of the series in Eq. (\ref{trans}) is guaranteed, 
because the photon number distributions of physical states normally decreases very 
quickly as $m$ increases, and the statistical erros also become small simply because 
we are dealing with diagonal elements of the density operator
in the number state basis \cite{ulf1}.
We need then to repeat this procedure $M$ times, for different
values of the driving field amplitude $\alpha_1,\; \alpha_2\; \dots\; \alpha_M$, 
to be able to cover enough points in phase-space and obtain the whole Wigner function 
of the original field.  

We have produced a numerical simulation of the above mentioned steps. Our simulation 
is going to be illustrated by the Wigner functions themselves. At $t=0.0$, a
Schr\"odinger cat state (Eq. \ref{scat}) is generated within the cavity. Its 
corresponding Wigner function is shown in Fig. 1, where we note the characteristic
interference structure. After the field has decayed, say, at $t=0.1/\gamma$, the loss
of coherence is indicated by the significant reduction of the interference structure, 
as seen in Fig. 2. Dissipation brings out the initially pure state very close to a 
statistical mixture \cite{vid}. We may instead drive the cavity with a coherent pulse 
(duration $t_d$), at $t=0$, 
in order to start our reconstruction procedure. After following the steps described above, we 
obtain the reconstructed Wigner function, as shown in Fig. 3, which is essentially the one 
in Fig. 1. We note that both peaks as well as the interference structure
characteristic of a Schr\"odinger cat state are entirely preserved. In our simulation
we have considered fluctuations that might be present during the measurement of
the atomic inversion. There might be experimental errors from various sources,
such as flucutuations in the amplitude of the driving field as well as in the
generated nonclassical state, which would cause distortions in the atomic inversion. 
In Fig. 4 we show the atomic inversion as a function of time and for a given value of 
the driving field amplitude $\beta=(0,2)$. We have obtained the field's Wigner function 
shown in Fig. 3 departing from a family of those ``distorted'' atomic inversions, for 
different values of $\beta$. 

\section{Conclusion}

In conclusion, we have presented a method for reconstructing the Wigner function of 
an initial nonclassical state at times when the field would have normally 
lost its quantum coherence. In particular, even at times such that the Wigner 
function would have lost its negativity, reflecting the decoherence process.
A crucial step in our approach is the driving of the initial
field immediately after preparation, which stores quantum coherences
in the diagonal elements of the time evolved displaced density
matrix, making them robust. We have therefore shown that the
initial displacement transfers the robustness of a coherent state \cite{zurek}
against dissipation to any initial state, allowing the full reconstruction of the
field state under less than ideal conditions. 

A natural application of our method would be the measurement of quantum states 
in cavities, where dissipation is difficult to avoid. Moreover, the application of the 
driving pulse at different times after the generation of a field state, would allow 
the ``snapshooting'' of the Wigner function as the state is dissipated. This means
that valuable information about the (mixed) quantum state as well as about the
decay mechanism itself could be retrieved while it 
suffers decay. The possibility of reconstructing quantum states even in 
the presence of dissipation may be also relevant for applications in quantum 
computing. Loss of coherence associated to dissipation is likely to occur in those 
devices, and our method could be used, for instance, as a scheme to refresh the 
state of a quantum computer \cite{ekert} in order to minimize the destructive action of 
the environment. 

\newpage

\begin{figure}[hp]
\vspace{0.3cm}
\centerline{\hspace{1.0cm}\psfig{figure=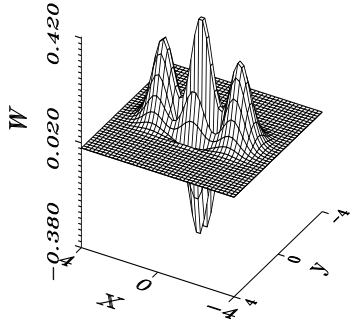,height=5cm,width=7cm}}
\vspace{0.5cm}
\caption{Wigner function of a Schr\"odinger cat state with $\alpha=2$ and $\phi=0$
at $t=0$.}
\end{figure}

\begin{figure}[hp]
\vspace{0.3cm}
\centerline{\hspace{1.0cm}\psfig{figure=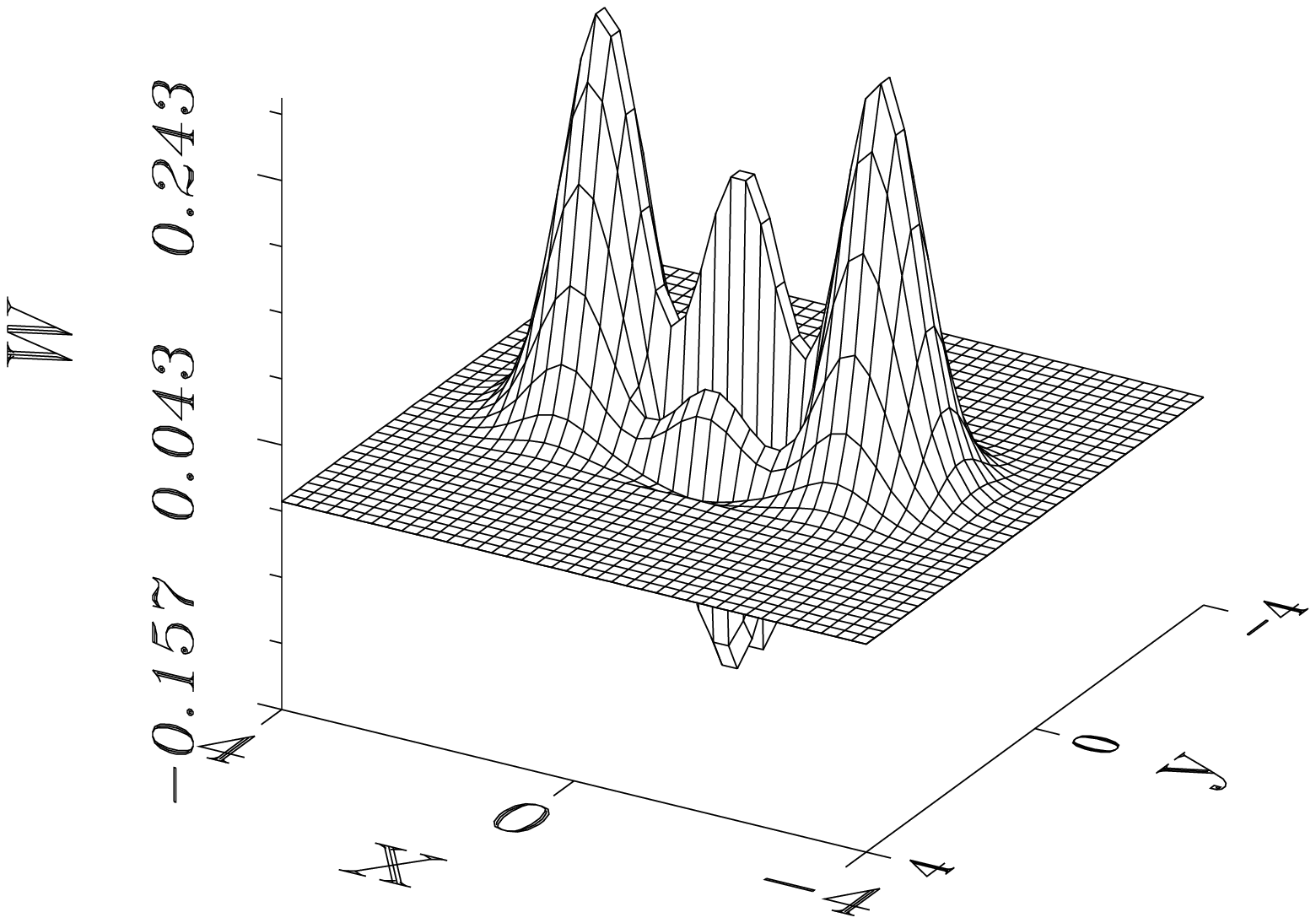,height=5cm,width=7cm}}
\vspace{0.5cm}
\caption{Wigner function of a Schr\"odinger cat state with $\alpha=2$ and $\phi=0$
at $t=0.1/\gamma$.}
\end{figure}

\begin{figure}[hp]
\vspace{0.3cm}
\centerline{\hspace{1.0cm}\psfig{figure=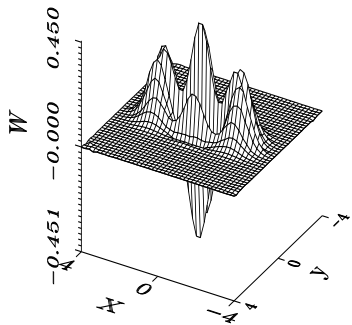,height=5cm,width=7cm}}
\vspace{0.5cm}
\caption{Wigner function of a Schr\"odinger cat state reconstructed at $t=0.1/\gamma$,
with $\alpha=2$ and $\phi=0$. The Wigner function at $t=0$ has been recovered 
(see Fig. 1).}
\end{figure}

\begin{figure}[hp]
\vspace{0.3cm}
\centerline{\hspace{1.0cm}\psfig{figure=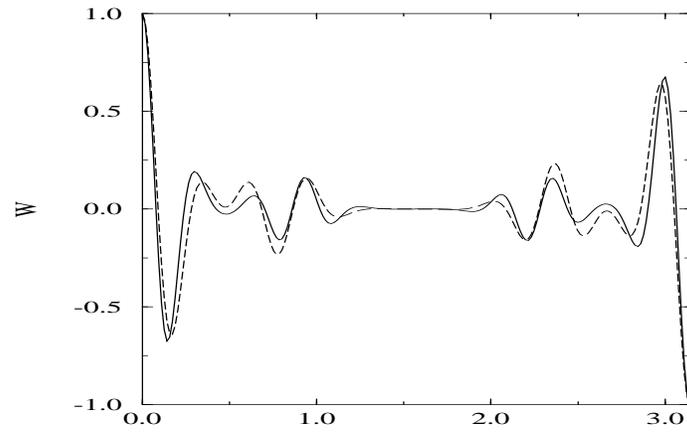,height=5cm,width=7cm}}
\vspace{0.5cm}
\caption{``Measured'' atomic inversion (dashed line) and ``exact'' atomic inversion 
of (continuous line) the displaced cavity field as a function of time for a displacement 
amplitude $\beta=(0,2)$.}
\end{figure}

\newpage

\acknowledgements

One of us, H.M.-C., thanks W. Vogel for useful comments.
This work was partially supported by FAPESP (Funda\c c\~ao de Amparo \`a
Pesquisa do Estado de S\~ao Paulo, Brazil), CONACYT (Consejo Nacional de
Ciencia y Tecnolog\'\i a, M\'exico), CNPq (Conselho Nacional de 
Desenvolvimento Cient\'\i fico e Tecnol\'ogico, Brazil) and ICTP (International
Centre for Theoretical Physics, Italy).

% \begin{figure}
% \caption{}
% \label{}
% \end{figure}

% \begin{table}
% \caption{}
% \label{}
% \end{tabular}
% \end{table}

\end{document}